\documentclass[aps,12pt,a4paper,superscriptaddress]{revtex4-1}
\usepackage{float}
\usepackage{amsmath}
\usepackage{amsfonts}
\usepackage{mathtools}
\usepackage{amssymb}
\usepackage{graphicx}
\usepackage{verbatim}
\usepackage{lipsum} 
\usepackage{float}
\usepackage{breqn}
 \usepackage{comment}
 \usepackage{color}
 \usepackage{hyperref}
 \usepackage{titlesec}
\usepackage[export]{adjustbox}
\usepackage{subfig}
\usepackage{breqn}
\date{\today}

\makeatletter
\let\cat@comma@active\@empty
\makeatother
\begin{document}

	\title{A Study of Berry Connection and Complex Analysis for Topological Characterization}

	\author{Y R Kartik}
	\author{ Rahul S}
	\author{Ranjith Kumar R}
		\affiliation{Poornaprajna Institute of Scientific Research, 4 Sadashivanagar Bangalore-560 080, India.}
		\affiliation{Manipal Academy of Higher Education, Madhava Nagar, Manipal, 576104, India.}
   	 \author{Sujit Sarkar}
   	 \affiliation{Poornaprajna Institute of Scientific Research, 4 Sadashivanagar Bangalore-560 080, India.}

\begin{abstract}%
\noindent We study and present the results of  Berry connection for the topological states in quantum matter. The Berry connection plays a central role in the geometric phase and topological phenomenon in quantum many-body system. We present the necessary and sufficient conditions to characterize the topological nature of the system through the complex analysis.
We also present the different topological aspects of the system in the momentum space.\\
\end{abstract}


\maketitle

	\section{Introduction}
	 	\noindent In nature, matter has different states such as solid, liquid and gaseous. These states are characterized by their internal structures. Other than internal structures, there exists symmetry. With discovery of fractional quantum Hall effect, there arose a new type of matter, called topological state of matter \cite{wen2016introduction}. These materials have a new type of internal order called topological order which makes them to be different from other materials. They are materials which conduct along the edge \cite{kane2005z} and are insulators at the bulk. The conduction along the edge is protected by two major symmetries \cite{chen2012symmetry, pollmann2012symmetry, chen2011classification}, time reversal and particle-conservation symmetry. The physics of Landau Fermi liquid theory provides a major description of the symmetry breaking phenomenon of quantum many-particle system. However the major limitations of Landau theory of phase transition is that it is related with the local order parameter. But it is well known for the study of the topological state of quantum matter that they do not have any order parameter.\\	 	
	 \noindent	Berry phase is the main tool to characterize the topological state of the system \cite{PhysRevLett.49.405,PhysRevLett.51.2167,hasan2010colloquium,rhim2017bulk,delplace2011zak,atala2013direct,zak1989berry}. 
	 When a time-dependent Hamiltonian with a state $|n\rangle  $ begins to evolve, it remains in the same state till the end. But along with the dynamical phase, it may also acquire a geometric phase depending on the path of evolution \cite{katanaev2012geometric,baggio2017aspects,chang2008berry,xiao2010berry}. For a quantum adiabatic process, geometric phase is generally known as Berry phase. The concept of Berry phase arises naturally from Berry connection  \cite{hanson2016notes,bohm2013geometric,chruscinski2012geometric,sato2017topological}. Berry connection is more fundamental than Berry phase and Berry curvature.
	It is non-observable and non-vanishing quantity which shows the overlapping of the wavefunction in an evolving system. It shows the connection from $\mathbb{R}$ to $\mathbb{R+dR}$ manifolds in the Hilbert space $\mathbb{H}$. In other-words it describes the parallel transport of the Block state in momentum space. There are observations of tensor Berry connection to explain the higher order gauge fields of quantum matter \cite{palumbo2019tensor}. By the proper selection of gauge one can explain the Berry connection of tensor monopoles found in 4D-Weyl type systems, Berry phase with or without Krammer degeneracy, monopoles of Dirac and Yang \cite{moore2017comment,Hatsugai_2010,PhysRevLett.121.170401}. To analyze the nature of topological state of matter, we need to understand the behavior of wavefunction.\\
	\noindent Topological properties can also be determined by the complex analysis of the parameter space \cite{chen2017elementary,SACRAMENTO2018216}. Topological phase transition points are considered as the singularity points in the parameter space. Topological phase transition results in the gapless condition of the energy spectrum. Topological invariants like winding number, Chern numbers are ill defined at these points. By using complex analysis method one can be define the topological properties both in gapless and gapped phases \cite{PhysRevLett.120.057001}. For the present study we show explicitly that how Berry connection is related to the topological quantum phase transition in the momentum space and we present a way of complex analysis to verify the method.\\
 
	\noindent\textbf{Motivation of the study:}	The main motivation of this work is to explain a few topological aspects of many-body system in an efficient manner.
\noindent Topological states of matter are characterized by the absence of local order parameters. But there are some studies which considers Berry phase as the local topological order parameter \cite{hatsugai2006quantized}. So, one can calculate the local topological order parameter (Berry phase) as a integration of Berry connection over the all available energy states. The Fourier transform of Berry connection is correlation function in 1D which gives the charge correlation function of Wannier states at different points \cite{hatsugai2004explicit,hatsugai2005characterization,hatsugai2006topological,fukui2005chern}. Recently there are studies to calculate correlation length, universality classes and scaling laws by using Berry connection \cite{chen2017correlation,chen2016scaling,molignini2018universal,Chen_2016}.\\
 One can characterize the topological state of a quantum system by the physical entity called Berry phase \cite{berry1985classical}. Berry phase exists where there is a closed interval. But Berry connection is a more fundamental quantity which reveals the path of evolution of ground-state wave function. Berry connection is a gauge dependent quantity which is equivalent to vector potential in electromagnetic theory \cite{hanson2016notes,gangaraj2017berry}. The significance of non-trivial Berry connection can be found in the Aharanov-Bohm effect \cite{girvin2019modern}. 
Here we are interested to show how one can directly study the topological behavior of the system just by studying the Berry connection and we argue that Berry connection is the fundamental tool to study the topological state of matter.\\ 
  The study of geometric phase to characterize the topological state of matter has already been done in the literature \cite{wilczek1989geometric}. It finally gives the parametric relation of the system to characterize the different topological aspects of the topological state of matter \cite{sarkar2018quantization}. But we show explicitly how the different topological aspects manifests in the momentum space. The other motivation of the study is to do the complex analysis to characterize the topological aspects of the problem. Complex analysis is a effective way to describe the state of a system. Depending on  the on the position of poles and zeroes, topological state of the system is determined \cite{ablowitz2003complex,SACRAMENTO2018216}. \\
  
\noindent\textbf{Outline of the work:}
\noindent In section \ref{Kit} we introduce and explain our model Hamiltonian.  
In section \ref{4} we present a detailed study of Berry connection and variation of topological angle. Here we give a clear picture between Berry connection, geometric phase and energy spectrum of different topological phases. In section \ref{2}, we explicitly show the existence of topological state from the perspective of complex analysis. Thus we verify how we can verify the topological properties of the system.\\ 

\section{Introduction to Model Hamiltonian}\label{Kit}
\noindent In this section, we first introduce the Hamiltonian of the present study. 
We consider the Kitaev chain as our model Hamiltonian \cite{kitaev2001unpaired,sarkar2018quantization,niu2012majorana}. It is a lattice model of p-wave superconductor in 1D \cite{kitaev2001unpaired}. Kitaev model has two phases. Topological phase for $\mu<2J$, non-topological phase for $\mu>2J$. Here the gapless condition occurs for the case $\mu=2J$ at $k=0$, where the gapless condition occurs. Through  energy dispersion study, we can understand gapless state formation \cite{sarkar2018quantization}. The Hamiltonian can be written as
		
\begin{equation}
H_0 = [\sum_{j}-{J} ({c_j}^{\dagger} c_{j+1} + h.c )-{\mu} {c_j}^{\dagger} {c_j}+{|\Delta|} ( {c_j} {c_{j+1}} + h.c ) ], \label{kitaev1}\end{equation}
where $ J$ is the hopping matrix element, $\mu$ is the chemical potential and $|\Delta |$ is the magnitude of the superconducting gap. We write the Hamiltonian in the momentum space as
\begin{dmath}
	H_1  =   \sum_{k> 0} ( \mu + 2J \cos k)
({\psi_k}^{\dagger} {\psi_k} + {\psi_{-k}}^{\dagger} {\psi_{-k}})
+  2i \Delta  \sum_{k > 0} \sin k  ({\psi_k}^{\dagger} {\psi_{-k}}^{\dagger} +
{\psi_{k}} {\psi_{-k}}),\end{dmath}
\noindent where $ {\psi^{\dagger}} (k) (\psi (k))$ is the creation (annihilation) operator of the spinless fermion of momentum $k$. 		
We can write the Hamiltonian in the BdG format as \begin{equation}
H_{BdG}(k)=\left(\begin{matrix}
\chi_1(k)&& i\chi_2(k)\\
-i\chi_2(k)&& -\chi_1(k)\\
\end{matrix}
\right).\end{equation}	
We can express the Hamiltonian by Anderson pseudo-spin approach \cite{anderson1958coherent,sarkar2017topological,niu2012majorana}. One can write the BdG Hamiltonian in the pseudo-spin basis as 
\begin{equation}
\vec{H}(k)=\chi_1(k)\vec{\tau_1}+\chi_2(k)\vec{\tau_2}+\chi_3(k) \vec{\tau_3}\Rightarrow H_{BdG}(k)=\Sigma_i\vec{\chi}_i(k).\vec{\tau}_i,\label{Kitaev}
\end{equation}\label{pseudo}

 \noindent where ${\tau_i}=(\tau_1,\tau_2,\tau_3)$ are the Pauli matrices, $\chi_1(k)=0$, $\chi_2(k)=2\Delta \sin k$ and $\chi_3(k)=-2J\cos k -\mu$.\\

 \noindent To characterize the topological trivial and non-trivial phases of the Kitaev chain one can construct the parametric space with axis, $X=\frac{\chi_1(k)}{|\chi(k)|}$ and $Y=\frac{\chi_2(k)}{|\chi(k)|}$ \cite{ablowitz2003complex}. The unit vector $\frac{\chi(k)}{|\chi(k)|}$ takes the closed path in the parametric space as $k$ varies across the Brillouin zone. Therefore the number of windings that the unit vector takes around its origin in the parametric space defines the winding number. The winding number $w=\pm n$ with $n=0,1,2,...$, characterizes the topological trivial ($w=0$) and non-trivial ($w=\pm 1$) phases in Kitaev chain.

	\section{Study of Berry connection and variation of topological angle for the model Hamiltonian}\label{4}
\noindent Here we present study of Berry connection and variation of topological angle within the Brillouin zone boundary. 
   The other important feature of the Berry connection is that it transforms as a vector gauge potential under gauge transformation. Therefore in quantum many-body lattice system, Berry connection associated with a Bloch state acts as an effective electrostatic vector potential defined in quantum metric space. Gauge invariant quantities can be defined from the Berry connection.\\
Here we mention very briefly how Berry connection is related with the Aharanov-Bohm effect to illustrate the importance of Berry connection.
 In this experimental set-up magnetic field $(\vec{B}=\vec{\nabla}\times\vec{A})$ outside the tube is zero \cite{wilczek1989geometric}. But the vector potential is non-zero. So for any closed trajectory around the flux tube, there is a global effect such that the line integral over the vector potential gives \cite{girvin2019modern}
\begin{equation}
\oint dr.\vec{A}(\vec{r})=n\Phi,
\end{equation}
where $\Phi$ is total flux inside the tube and
$n$ is integer winding number of the trajectory around the tube. It has shown explicitly in \cite{sarkar2018quantization} that the phase acquired in Aharanov-Bohm effect is the topological phase. Therefore the non-local effect of $\bar{A}$ causes the topological phase. By using Stokes theorem, one can connect the Berry phase and Berry connection through the analytical relation, $\gamma_n=\oint_c A_n(r).dr$, over the closed contour $c$. The resulting Berry phase will be $2\pi$ or integral multiples of $2\pi$. If the contour is not closed Berry phase won't exists, because gauge dependency is not present.
 
\noindent The geometric phase acquired by the the wavefunction in Aharanov-Bohm effect is a gauge-independent quantity. The phase acquired by the wavefunction is
\begin{equation}
\phi_k=\int_{s}F(k).ds,
\end{equation}
where $F(k)$ which is nothing other than the magnetic flux density or Berry curvature. It is a gauge independent quantity. It can also be written as
\begin{equation}
F(k)=\nabla_k\times A(k)=\nabla_k\times \langle \psi(k)|i\nabla_k|\psi(k)\rangle 
=i\langle\nabla_k\psi(k)|\times|\nabla_k\psi(k)\rangle
\end{equation}
Where $A(k)$ is the Berry connection. Berry curvature is a geometric property of the system. It can be observes as magnetic field effects in electronic as well as optical Hall effects. Sometimes there arises the singularities in the parameter space of the vector potential. In such cases Stokes theorem becomes invalid and the geometric phase is ill defined. Otherwise geometric phase is always the integral multiple of $2\pi$.
\begin{equation}
\phi_k=\int_{s}F(k).ds=\oint_cA(k).dl=\gamma
\end{equation}

\noindent It is clear from the exclusive analysis that the Berry phase which is analogous to the magnetic flux may be zero but Berry connection which is analogous to the vector potential is always non-zero. The important result of this study is that for the topological state of matter like topological insulator and topological superconductors, this relation between the Berry connection and Berry phase is verified.

\noindent Mathematically for our model Hamiltonian Berry connection is expressed as $A_k=\left\langle \psi_{k}|i\partial_k|\psi_{k}\right\rangle.$
From eq. \ref{Kitaev}, we get the eigenvalues, 
\begin{equation}
E_k=\pm\sqrt{(2J\cos k+\mu)^2+(2\Delta\sin k)^2}=\sqrt{2(J^2+\Delta^2)+\mu^2+2J\mu\cos k+2(j^2-\Delta^2)\cos 2k}.
\end{equation}

\noindent Eigenvectors are given by 
\begin{dmath}
|\psi\rangle=-\frac{i}{2\Delta}(\mu+2J\cos k-E_k)\csc k|\uparrow\rangle+1|\downarrow\rangle,
\end{dmath}
where, $|\uparrow\rangle =(1,0)^T$ and $|\downarrow\rangle =(0,1)^T$ are the column matrices.\\
Berry connection is given by
\begin{dmath}
A_k=\left\langle \psi_{k}|i\partial_k|\psi_{k}\right\rangle=i\left(-\frac{i}{2\Delta}(\mu+2J\cos k+E_k\csc k)\right)
*\left(\frac{i}{2\Delta}(\mu+2J\cos k+E_k\csc k)\\-i\csc k(-2J\sin k+\frac{-4J\mu \sin k-4(J^2-\Delta^2)\sin 2k}{E_k})\right).
\end{dmath}
After simplification we get
\begin{dmath}
A_k=-i\frac{(2J+\mu\cos k)(\mu+2J\cos k+E_k)^2\csc^3 k}{4\Delta^2E_k}.
\end{dmath}

\begin{figure}[H]
 	\centering
 	\includegraphics[width=9cm,height=10cm]{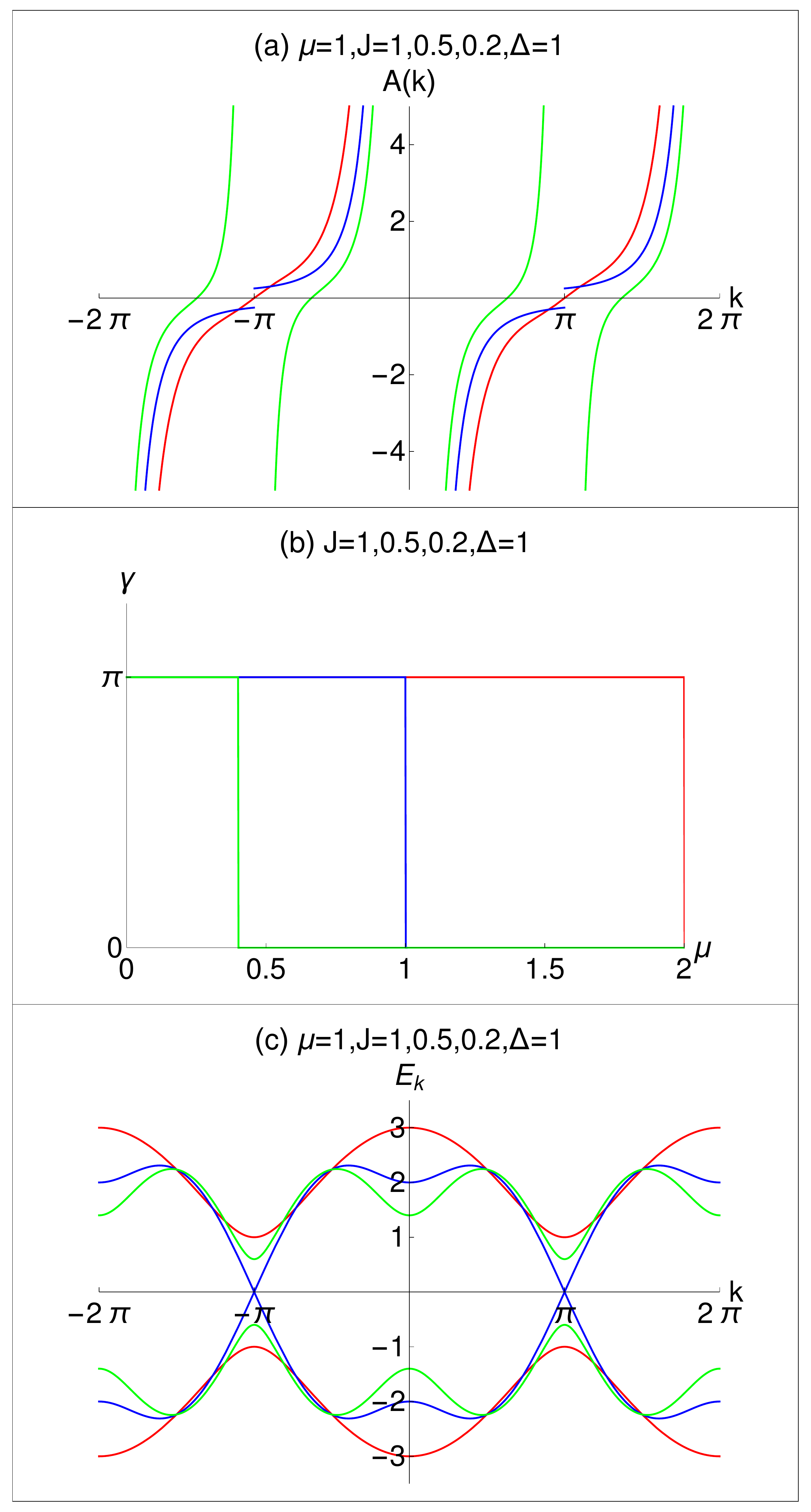}
 	\caption{The upper panel of the figure shows the variation of Berry connection for Kitaev Hamiltonian ($H_k$) with $k$. The red curve represents the topological case($\mu<2J$), the blue curve represents transition state ($\mu=2J$) and the green curve represents non-topological state ($\mu>2J$). The middle panel shows the geometric phase for corresponding parameter spaces. The lower panel shows the energy dispersion curves for the corresponding parametric spaces.}
 	\label{ac1}
 \end{figure}

\noindent Fig. \ref{ac1} contains three panels. The upper panel shows the variation of $A(k)$ with momentum $k$. This figure presents three curves for different values of $J$  ($J=1,0.5,0.2$). These different curves are for different topological states of system. i.e. $\mu<2J$,  $\mu=2J$ and  $\mu>2J$, as one can find from the topological invariant number \cite{sarkar2017topological}. It is well known in the literature that at the topological quantum phase transition point $(\mu=2J)$ winding number shows the discontinuous value. In the present study we show explicitly at this particular point, at the first Brillouin zone boundary there is also a discontinuity in the Berry connection. We show that at the topological quantum phase transition point, there is a jump of $A(k)$. For the topological states the $A(k)$ curve is continuous over the BZ boundary and for the non-topological states $A(k)$ is within the BZ boundary.
The middle panel shows the energy dispersion relations with $k$ for the corresponding three phases. One can observe how the energy gap closes for the topological quantum phase transition point. But for the topological and non-topological phases, there is a gap in BZ. The system goes from one topological state to the other by closing the gap. \\

\begin{figure}[H]
 	\centering
 	\includegraphics[width=9cm,height=10cm]{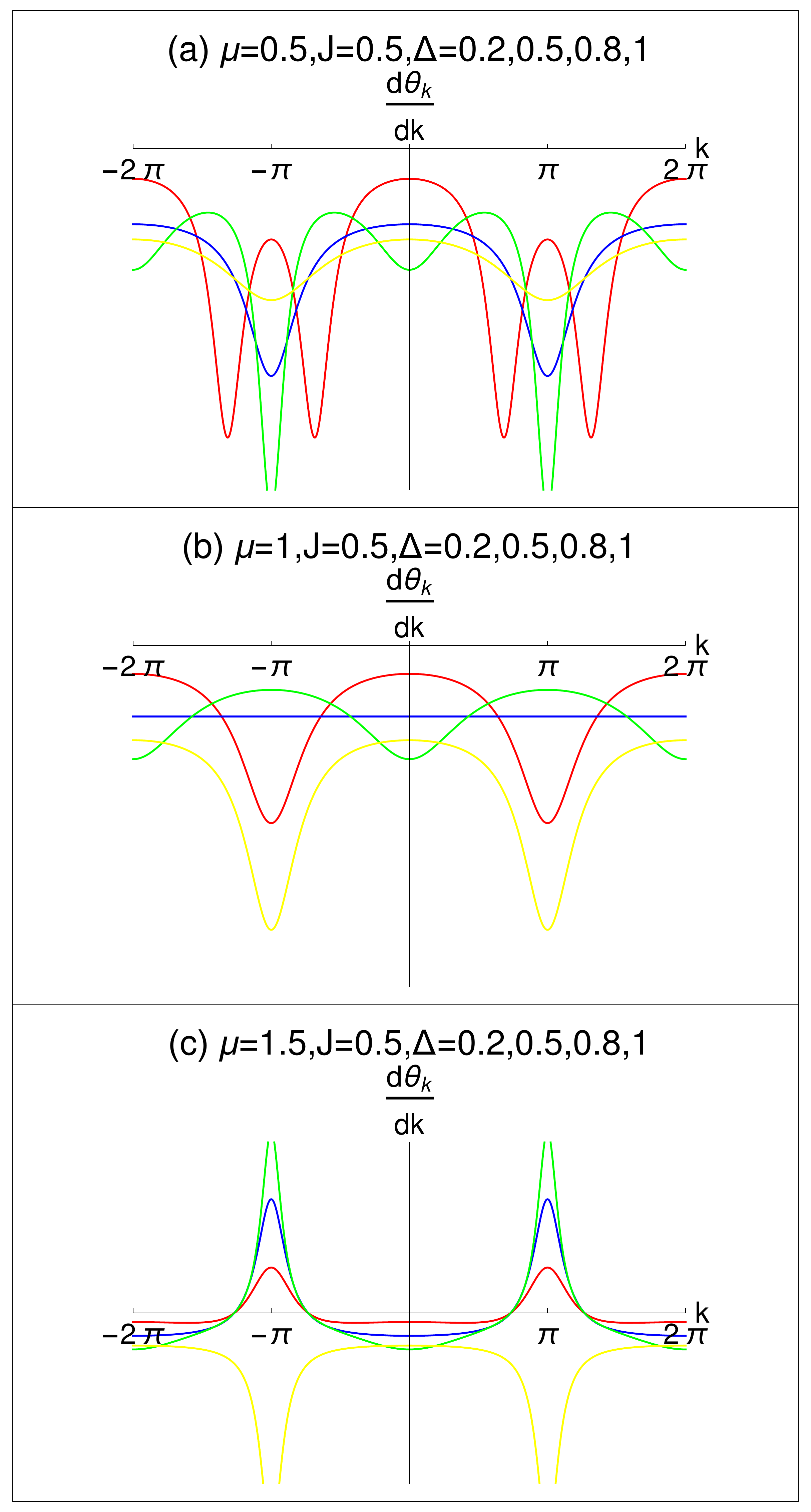}
 	\caption{Variation of topological angle $(d\theta_k/dk)$ for Kitaev Hamiltonian (H) with $k$. The upper panel shows topological case($\mu<2J$), middle panel shows transition case ($\mu=2J$) and the lower panel shows non-topological case ($\mu>2J$). Red, blue, green and yellow curves represent $\Delta=0.2,0.5,0.8$ and 1 respectively}
 	\label{ta}
 \end{figure}

\noindent In the rotated basis we can write the Hamiltonian as
 \begin{equation}
 H(k) =  (\epsilon_{k}- \mu) \sigma_{x} - 2\Delta\sin k \sigma_{y},
 \end{equation}
 where $\epsilon_{k} = -2t\cos k$. We write the Hamiltonian in matrix form as

 $$ H= \begin{bmatrix}
 0  && r e^{i\theta} \\
 r e^{-i\theta} && 0 
 \end{bmatrix}, $$ 
 
 \noindent where  $ r = \sqrt{(2t\cos k+\mu)^2 + 4\Delta^2 \sin^2 k}$\;\; and \;\;$ \theta_k = -\tan^{-1}(\frac{2\Delta\sin k}{2t\cos k+\mu}).$\\
$\theta_k$ is topological angle, which is the angle made by wave-vector in the Brillouin zone. The integration over the variation of topological angle with wave-vector gives the geometric phase. i.e.
\begin{equation}
\gamma=\int_{-\pi}^{+\pi}\left(\frac{d\theta_k}{dk}\right)dk.
\end{equation} 
So, present model $\left(\frac{d\theta_k}{dk}\right)$ is given by
\begin{equation}
\frac{d\theta_k}{dk}=\frac{2\Delta(2J+\mu\cos k)}{(2J\cos k+\mu)^2+(2\Delta\sin k)^2}.
\end{equation}\label{dt}
 \noindent Fig. \ref{ta} shows the variation of the topological angle $\left(\frac{d\theta_k}{dk}\right)$ with the momenta $k$. The figure consists of three panels. Upper panel is for topological $(\mu<2J)$, middle is for critical $(\mu=2J)$, and lower is for non-topological $(\mu>2J)$ case respectively. Each figure consists of four curves for different values of $\Delta$ ($\Delta=0.2,0.5,0.8,1$). 
The variation of $\theta_k$ is almost absent for non-topological state, except at the very edge of the Brillouin zone boundary. We observe that the variation of topological angle is maximum for topological state of matter. We observe that variation is more prominent for lower values of $\Delta$. We also observe that as the system goes from the topological state to the non-topological state, the variation of $\theta_k$ is small. This behavior of $\theta_k$ in momentum space is physically consistent with the behavior of winding number.\\

\noindent This study of $\frac{d\theta_k}{dk}$ show a reflection of topology in the momentum space. i.e. how the karnel of eq. \ref{dt} behaves in the momentum space. It reveals for this study that the variation of topological angle for different topological states are different. 
\section{An analysis of topology of the model Hamiltonian from the perspective of complex variable}\label{2}

      \noindent The argument principle of complex analysis allows one to calculate winding number of simple closed contour about its origin. Argument principle states that the winding number can be written as a contour integral of a meromorphic function $f(z)$. This integral can be expressed in terms of number of zeros and poles of $f(z)$. Therefore the winding number $w$ of a simple closed contour about its origin, can also be expressed as a difference between number of zeros and poles of $f(z)$, i.e. $w=Z-P$ \cite{ablowitz2003complex}, where $Z$ and $P$ are zeros and poles respectively.\\
\noindent The generic form of the Hamiltonian can be written as 
\begin{equation}
H(k)=d(k).\sigma= \left( \begin{matrix}
0 && q^*(k)\\
q(k) && 0 
\end{matrix}\right),
\end{equation}
where $q(k)=-2J\cos k-\mu+2i\Delta \sin k$. The winding number can be calculated using the definition \cite{sacramento2018duality}
\begin{equation}
w=\frac{1}{2\pi i}\int\limits_{-\pi}^{\pi} dk \frac{\partial_k q(k)}{q(k)}
= \frac{1}{2\pi i}\int\limits_{-\pi}^{\pi} dk \frac{2J\sin k + 2i\Delta \cos k}{-2J \cos k -\mu + 2i\Delta \sin k}\\ \label{C1}
\hspace{0.3cm}.
\end{equation}
Writing $\sin k$ and $\cos k$ in the exponential form and substituting $z=e^{ik}$ and $dz=ie^{ik}dk$, one can rewrite the above equation \cite{ablowitz2003complex,d2010introduction} as 
\begin{equation}
w=\frac{1}{2\pi i}\int\limits_{-\pi}^{\pi} \frac{dz}{z} \frac{\Delta(z^2+1)-t(z^2-1)}{(\Delta-t)z^2 - \mu z - (\Delta+t)}
\hspace{0.3cm}.\end{equation}

\begin{figure}[H]
	\centering
	\includegraphics[width=14cm,height=4cm]{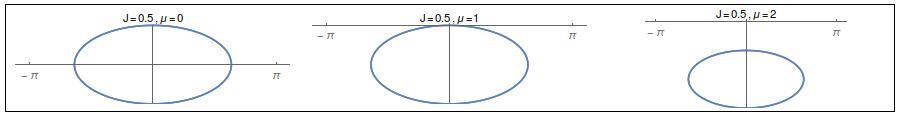}
	\caption{Parametric plot for the Hamiltonian $H(k)$ for different values of $\mu$.} \label{aux1}
\end{figure}

\begin{figure}[H]
 	\centering
 	\includegraphics[width=10cm,height=8cm]{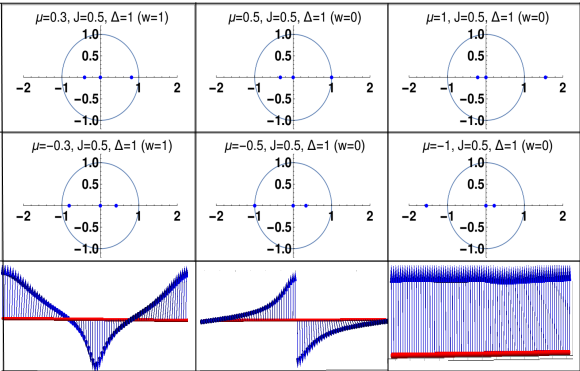}
 	\caption{Zeros and poles of the complex function $f(z)$. Left panel: topological phase, middle panel: phase transition case and right panel: non-topological phase respectively.}
 	\label{comp}
 \end{figure}

\noindent Winding number can be calculated using the argument principle of complex analysis. eq. \ref{C1} has a general form as
\begin{equation}
\frac{1}{2\pi i} \oint \frac{f^{\prime}(z)}{f(z)} dz.
\end{equation}
This integral can be solved by calculating number of poles ($P$) and number of zeros ($Z$) of the function $f(z)$ as
\begin{equation}
\frac{1}{2\pi i} \oint \frac{f^{\prime}(z)}{f(z)} dz = Z-P.
\end{equation}
In our case $f(z)$ has the form $f(z)= \frac{1}{z}\left( \Delta(z^2-1)-J(z^2+1)-\mu z \right) $. Clearly it has a pole at $z=0$. It has two zeros, one at $z=+1$ if $\mu=-2J$, another at $z=-1$ if $\mu=2J$. Therefore the winding number is $w=2-1=1$ if $-2J<\mu<2J$. Thus, in parametric space it encircles the origin one time. Either the winding number is $+1$ or $-1$. The sign indicates the direction of encircling, clockwise or counterclockwise. This corresponds to the topological region of the Hamiltonian $H(k)$. If there are equal numbers of poles as well as zeros, there will be no winding number. Here the curve will not encircle the origin. This shows non-topological case. There are some cases when the zeros and poles are present on the contour of the curve. At this point we can not explain the winding number for the particular function. Here the curve touches the origin and the function is ill defined. This point is known as topological phase transition point (fig.\ref{aux1}). \\
\noindent Fig \ref{aux1} contains three panels. Left one indicates the topological state of the system where the origin is encircles by a closed curve. Middle panel represents the topological phase-transition case where the curve touches the origin. Right panel shows the non-topological state of the of the system where origin lies outside the closed curve. This is a straight forward method to analyze the topological properties of the system.\\   
\noindent The same results can be obtained in slightly different approach. First we write the Hamiltonian (eq.1) in Majorana operators with $c_n^{\dagger}=\frac{a+ib}{2}$ and $c_n=\frac{a-ib}{2}$,
\begin{equation}
H=-i\left[\sum\limits_{j=1}^{N-1} \left( \frac{J-\Delta}{2}\right)b_ja_{j-1} - \sum\limits_{j=1}^{N}\left( \frac{\mu}{2}\right)b_ja_j + \sum\limits_{j=1}^{N-1} \left( \frac{J+\Delta}{2}\right)b_ja_{j+1}\right] 
\end{equation}
Generic form of this Hamiltonian can be written as
\begin{equation}
H=\sum_{\alpha=-1,0,1} \gamma_{\alpha}H_{\alpha},
\end{equation}
where $H_{\alpha}=\sum\limits_{j=1}^{N}b_ja_{j+\alpha}$ and $\gamma_{\alpha=-1,0,1}=\left( \frac{J-\Delta}{2}\right), \mu, \left( \frac{J+\Delta}{2}\right)$ respectively. The Fourier transform of the Hamiltonian is $f(k)=\sum_{\alpha=-1,0,1}\gamma_{\alpha}e^{ik\alpha}$. With $z=e^{ik}$ we get the complex function associated with the Hamiltonian,
\begin{equation}
f(z)= \sum_{\alpha=-1,0,1}\gamma_{\alpha}z^{\alpha}= \frac{1}{z}\gamma_{-1}-\gamma_0+z\gamma_1= \frac{1}{z}\left(\gamma_{1}z^2-\mu z+\gamma_{-1} \right) 
\end{equation}
Thus we have the roots $z_{1,2}=\frac{\mu\pm\sqrt{\mu^2 - 4\gamma_1\gamma_{-1}}}{2\gamma_1}=\frac{\mu\pm\sqrt{\mu^2-J^2+\Delta^2}}{J+\Delta}$. The complex function $f(z)$ has a pole at $z=0$ and two zeros at $z=z_1$ and $z=z_2$. Topological, non-topological phases and the transition between them can be characterized by observing whether two zeros fall inside, outside or on the unit circle respectively. Using the argument principle of complex analysis, topological invariant $w=Z-P$ (where $Z$ is number zeros that falls inside the unit circle and $P$ is number of pole) can be calculated. In our case the pole is always at the origin of the unit circle. But values of two zeros $z_{1,2}$ depends on the parameter values. We already know for parameter condition $-J<\mu<J$ the system is in the topological phase and for $\mu<-J$ and $\mu>J$, system is in non-topological phase. $\mu=\pm J$ is thus the gap closing point indicating the topological quantum phase transition. This is shown more explicitly in the fig.\ref{comp} in terms number of zeros and pole of $f(z)$. Left panel of fig \ref{comp} shows topological phase for finite $\mu$ satisfying $-J<\mu<J$. We observe both the zeros lies inside the unit circle giving $w=2-1=1$. The middle and right panel of fig \ref{comp} shows the system at the phase transition point and at non-topological phase respectively. At $\mu=\pm J$ we see one of the zero lie on the unit circle while another lie inside, which gives $w=0$. For non-topological phase one of the zero lie inside and another lie outside the unit circle giving $w=0$.

\noindent Figure \ref{comp} consists of three panels. Left panel represent the topological case. Here we can observe both zeros lies inside the unit circle. The corresponding pseudo-spin (eq. \ref{pseudo}) vector encircles the origin without any discontinuity. The middle panel represents the topological phase-transition case. Here the zeros lies on the unit circle. The corresponding pseudo- spin vector encircles the origin but there is a discontinuity at the phase-transition point. The right panel represents the non-topological case. Here the zeros lies outside the unit circle. The corresponding pseudo-spin vector does not encircles the origin.



	\noindent\textbf{Conclusion:}
We presented the detailed analysis of the Berry connection for the Kitaev model Hamiltonian and explained how the Berry connection can be helpful for the understanding of the topological properties of the system. We showed that how the topological angle varies in the momentum-space as we vary the parameters of the system. We gave the explicit explanation for the existence of topological state through the complex analysis technique. \\\\ 
\noindent\textbf{Acknowledgments:}
    The authors would like to acknowledge DST (EMR/2017/000898) for the funding and RRI library for the books and journals. The authors would like to acknowledge Mr. N. A. Prakash for reading this manuscript critically and Mr. Amitava Banerjee for useful discussions. Finally authors would like to acknowledge ICTS lectures/seminars/workshops/conferences/discussion meetings on different aspects of physics.\\ 


%

\bibliography{CurvatureNotesEPL}

\begin{thebibliography}{10}

\bibitem{wen2016introduction}
Xiao-Gang Wen.
\newblock {\em URL http://web. mit.
  edu/physics/people/faculty/docs/wen\_intro\_topological\_orders. pdf}, 34,
  2016.

\bibitem{kane2005z}
Charles~L Kane and Eugene~J Mele.
\newblock {\em Physical review letters}, 95(14), 2005.

\bibitem{chen2012symmetry}
Xie Chen, Zheng-Cheng Gu, Zheng-Xin Liu, and Xiao-Gang Wen.
\newblock {\em Science}, 338(6114), 2012.

\bibitem{pollmann2012symmetry}
Frank Pollmann, Erez Berg, Ari~M Turner, and Masaki Oshikawa.
\newblock {\em Physical review b}, 85(7), 2012.

\bibitem{chen2011classification}
Xie Chen, Zheng-Cheng Gu, and Xiao-Gang Wen.
\newblock {\em Physical review b}, 83(3), 2011.

\bibitem{PhysRevLett.49.405}
D.~J. Thouless, M.~Kohmoto, M.~P. Nightingale, and M.~den Nijs.
\newblock {\em Phys. Rev. Lett.}, 49:405--408, 1982.

\bibitem{PhysRevLett.51.2167}
Barry Simon.
\newblock {\em Phys. Rev. Lett.}, 51:2167--2170, 1983.

\bibitem{hasan2010colloquium}
M~Zahid Hasan and Charles~L Kane.
\newblock {\em Reviews of Modern Physics}, 82(4), 2010.

\bibitem{rhim2017bulk}
Jun-Won Rhim, Jan Behrends, and Jens~H Bardarson.
\newblock {\em Physical Review B}, 95(3), 2017.

\bibitem{delplace2011zak}
Pierre Delplace, D~Ullmo, and G~Montambaux.
\newblock {\em Physical Review B}, 84(19), 2011.

\bibitem{atala2013direct}
Marcos Atala, Monika Aidelsburger, Julio~T Barreiro, Dmitry Abanin, Takuya
  Kitagawa, Eugene Demler, and Immanuel Bloch.
\newblock {\em Nature Physics}, 9(12), 2013.

\bibitem{zak1989berry}
J~Zak.
\newblock {\em Physical review letters}, 62(23), 1989.

\bibitem{katanaev2012geometric}
MO~Katanaev.
\newblock {\em Russian Physics Journal}, 54(10), 2012.

\bibitem{baggio2017aspects}
Marco Baggio, Vasilis Niarchos, and Kyriakos Papadodimas.
\newblock {\em Journal of high energy physics}, 2017(4), 2017.

\bibitem{chang2008berry}
Ming-Che Chang and Qian Niu.
\newblock {\em Journal of Physics: Condensed Matter}, 20(19), 2008.

\bibitem{xiao2010berry}
Di~Xiao, Ming-Che Chang, and Qian Niu.
\newblock {\em Reviews of modern physics}, 82(3), 2010.

\bibitem{hanson2016notes}
George~W Hanson, S~Gangaraj, and Andrei Nemilentsau.
\newblock {\em arXiv preprint arXiv:1602.02425}, 2016.

\bibitem{bohm2013geometric}
Arno Bohm, Ali Mostafazadeh, Hiroyasu Koizumi, Qian Niu, and Josef Zwanziger.
\newblock {\em The Geometric Phase in Quantum Systems: Foundations,
  Mathematical Concepts, and Applications in Molecular and Condensed Matter
  Physics}.
\newblock Springer Science \& Business Media, 2013.

\bibitem{chruscinski2012geometric}
Dariusz Chruscinski and Andrzej Jamiolkowski.
\newblock {\em Geometric phases in classical and quantum mechanics}, volume~36.
\newblock Springer Science \& Business Media, 2012.

\bibitem{sato2017topological}
Masatoshi Sato and Yoichi Ando.
\newblock {\em Reports on Progress in Physics}, 80(7), 2017.

\bibitem{palumbo2019tensor}
Giandomenico Palumbo and Nathan Goldman.
\newblock Tensor berry connections and their topological invariants.
\newblock {\em Physical Review B}, 99(4):045154, 2019.

\bibitem{moore2017comment}
Gregory~W Moore.
\newblock A comment on berry connections.
\newblock {\em arXiv preprint arXiv:1706.01149}, 2017.

\bibitem{Hatsugai_2010}
Yasuhiro Hatsugai.
\newblock {\em New Journal of Physics}, 12(6):065004.

\bibitem{PhysRevLett.121.170401}
Giandomenico Palumbo and Nathan Goldman.
\newblock {\em Phys. Rev. Lett.}, 121:170401, 2018.

\bibitem{chen2017elementary}
Bo-Hung Chen and Dah-Wei Chiou.
\newblock {\em arXiv preprint arXiv:1705.06913}, 2017.

\bibitem{SACRAMENTO2018216}
P.D. Sacramento and V.R. Vieira.
\newblock Duality and topology.
\newblock {\em Annals of Physics}, 391:216 -- 239, 2018.

\bibitem{PhysRevLett.120.057001}
Ruben Verresen, Nick~G. Jones, and Frank Pollmann.
\newblock {\em Phys. Rev. Lett.}, 120:057001, 2018.

\bibitem{hatsugai2006quantized}
Yasuhiro Hatsugai.
\newblock {\em Journal of the Physical Society of Japan},
  75(12):123601--123601, 2006.

\bibitem{hatsugai2004explicit}
Yasuhiro Hatsugai.
\newblock {\em Journal of the Physical Society of Japan}, 73(10):2604--2607,
  2004.

\bibitem{hatsugai2005characterization}
Yasuhiro Hatsugai.
\newblock {\em Journal of the Physical Society of Japan}, 74(5):1374--1377,
  2005.

\bibitem{hatsugai2006topological}
Y~Hatsugai, T~Fukui, and Hiroshi Suzuki.
\newblock {\em Physica E: Low-dimensional Systems and Nanostructures},
  34(1-2):336--339, 2006.

\bibitem{fukui2005chern}
Takahiro Fukui, Yasuhiro Hatsugai, and Hiroshi Suzuki.
\newblock {\em Journal of the Physical Society of Japan}, 74(6):1674--1677,
  2005.

\bibitem{chen2017correlation}
Wei Chen, Markus Legner, Andreas R{\"u}egg, and Manfred Sigrist.
\newblock {\em Physical Review B}, 95(7):075116, 2017.

\bibitem{chen2016scaling}
Wei Chen, Manfred Sigrist, and Andreas~P Schnyder.
\newblock {\em Journal of Physics: Condensed Matter}, 28(36):365501, 2016.

\bibitem{molignini2018universal}
Paolo Molignini, Wei Chen, and Ramasubramanian Chitra.
\newblock {\em Physical Review B}, 98(12):125129, 2018.

\bibitem{Chen_2016}
Wei Chen, Manfred Sigrist, and Andreas~P Schnyder.
\newblock {\em Journal of Physics: Condensed Matter}, 28(36):365501, 2016.

\bibitem{berry1985classical}
MV~Berry.
\newblock {\em Journal of physics A: Mathematical and general}, 18(1):15, 1985.

\bibitem{gangaraj2017berry}
S~Ali~Hassani Gangaraj, M{\'a}rio~G Silveirinha, and George~W Hanson.
\newblock {\em IEEE journal on multiscale and multiphysics computational
  techniques}, 2:3--17, 2017.

\bibitem{girvin2019modern}
Steven~M Girvin and Kun Yang.
\newblock {\em Modern condensed matter physics}.
\newblock Cambridge University Press, 2019.

\bibitem{wilczek1989geometric}
Frank Wilczek and Alfred Shapere.
\newblock {\em Geometric phases in physics}, volume~5.
\newblock World Scientific, 1989.

\bibitem{sarkar2018quantization}
Sujit Sarkar.
\newblock {\em Scientific reports}, 8(1):5864, 2018.

\bibitem{ablowitz2003complex}
Mark~J Ablowitz and Athanassios~S Fokas.
\newblock {\em Complex variables: introduction and applications}.
\newblock Cambridge University Press, 2003.

\bibitem{kitaev2001unpaired}
A~Yu Kitaev.
\newblock {\em Physics-Uspekhi}, 44(10S):131, 2001.

\bibitem{niu2012majorana}
Yuezhen Niu, Suk~Bum Chung, Chen-Hsuan Hsu, Ipsita Mandal, S~Raghu, and Sudip
  Chakravarty.
\newblock {\em Physical Review B}, 85(3):035110, 2012.

\bibitem{anderson1958coherent}
Philip~W Anderson.
\newblock {\em Physical review}, 110(4), 1958.

\bibitem{sarkar2017topological}
Sujit Sarkar.
\newblock {\em Scientific Reports}, 7(1), 2017.

\bibitem{sacramento2018duality}
PD~Sacramento and VR~Vieira.
\newblock {\em Annals of Physics}, 391, 2018.

\bibitem{d2010introduction}
John~P D'Angelo.
\newblock volume~12.
\newblock American Mathematical Soc., 2010.

\end{thebibliography}

\end{document}